\documentstyle[preprint,aps]{revtex}
 
\begin{document}
\draft
\tightenlines

\title{Self-organized criticality as an absorbing-state phase transition}
\author{Ronald Dickman$^{1,*}$, Alessandro Vespignani$^{2,\dagger}$ and Stefano Zapperi$^{3,\ddagger}$}
\address{$^1$Department of Physics and Astronomy, Lehman College, CUNY,
Bronx, NY 10468-1589 \\
$^2$International Centre for Theoretical Physics (ICTP) 
P.O. Box 586, 34100 Trieste, Italy\\
$^3$Center for Polymer Studies and Department of Physics,
Boston University, Boston, MA 02215}

\date{\today}

\maketitle
\begin{abstract}

We explore the connection between 
self-organized criticality and phase transitions in models with absorbing states.
Sandpile models are found to exhibit criticality only when a pair of relevant
parameters --- dissipation $\epsilon$ and driving field $h$ --- are set to their
critical values.  The critical values of $\epsilon$ and $h$ are both 
equal to zero. The first result is due to the absence of saturation 
(no bound on energy) in the sandpile
model, while the second result is common to other 
absorbing-state transitions. The original definition of the
sandpile model places it at the point ($\epsilon=0,h=0^+$): it is 
{\em critical by definition}.  We argue power-law avalanche 
distributions are a general feature of models with
infinitely many absorbing configurations, when they are subject to slow
driving at the critical point. Our assertions are supported by simulations
of the sandpile at $\epsilon=h=0$ and {\em fixed} energy density
$\zeta$ (no drive, periodic
boundaries), and of the slowly-driven pair contact process.  
We formulate a field theory for the sandpile model,
in which the order parameter is coupled to
a {\em conserved} energy density, which plays the role of an effective
creation rate.

\end{abstract}


\section{Introduction}

Avalanche behavior is common to many physical phenomena,
ranging from magnetic systems (the Barkhausen effect)
\cite{bark} and flux lines in high-$T_c$ superconductors \cite{flux},
to fluid flow through porous media \cite{porous},
microfracturing processes \cite{ae},
earthquakes \cite{gr}, and lung inflation \cite{lung}.
The common feature of all these systems is slow 
external driving, causing an intermittent,  
widely distributed response. 
Avalanches come in very different sizes, often
distributed as a power law.
This fact excites the interest of statistical physicists, since
power laws imply the absence of a characteristic scale,
a feature observed close to a critical point. 
In order to describe a critical point, we need only
specify a set of critical exponents, whose values are determined
by general symmetries and conservation laws and do not depend on
microscopic details of the system. 

Is there a connection between the observed 
power-law distribution of avalanche sizes and critical phenomena? 
And if so, can we understand the physics of
avalanches by applying what we know about critical points
and universality?
A tentative answer to these questions was given by 
Bak, Tang and Wiesenfield (BTW) \cite{btw}, who proposed that
the power laws in avalanche statistics are due to
a new kind of critical phenomenon, which they called self-organized criticality
(SOC). In ordinary phase transitions, criticality is
attained only by fine-tuning certain control parameters
(temperature, pressure, etc.) to special values.
Only close to this critical point is scale invariance observed.
BTW suggested that in some dynamical systems
the critical point is reached automatically, without any fine
tuning, thus explaining the wide occurrence of power laws
in nature. The idea was then exemplified by several
dynamical models, such as the sandpile \cite{btw}, and the forest-fire
model\cite{ff}.  The SOC hypothesis has
stimulated an enormous amount of research.
If this --- as one of its originators has concluded --- is
``how nature works'' \cite{how}, then the question of
``how SOC works'' becomes all the more urgent.

The concept of ``spontaneous'' criticality, as 
discussed in the SOC literature, presents, however, several
ambiguities. Several authors have noted
that the external driving rate is a parameter that has to be 
fine-tuned to zero in order to observe criticality \cite{grin,sor1,ddrg,vz}. 
On the other hand, it would be amazing, without prior knowledge of the critical
coupling, to define a system like the Ising model
so that it is intrinsically at its critical point.
SOC appears less miraculous if we suppose that there is `generic
scale invariance,' i.e., that criticality obtains over a region of parameter
space not just a point.  But we will argue that as for the Ising model, 
SOC typically exists at a critical {\em point} in the relevant parameter space. 
The identity of the parameters has been obscured by the manner in which
the models were defined.  How can a model be critical by definition, 
when for most statistical mechanics models, we don't even 
{\em know} the exact critical point?
One way for a system to discover its own critical point is through a suitable
extremal dynamics, as in invasion percolation; another is that we may know the
critical parameters {\em a priori}, because they are fixed by a symmetry or a
conservation law, and build these into the definition of the model.  

Recently, a novel mean-field analysis of SOC models was presented \cite{vz},
which pointed out the similarities between SOC models and
models with absorbing states \cite{RDPRV,marro}.  (An absorbing state 
is one allowing no further change or activity.)
The mean-field theory provides a new
insight into the origin of SOC, which, in the sandpile model,
is essentially the criticality of the population of toppling sites.
It turns out that SOC
corresponds to the onset of non-locality in the dynamics of the system.
Non-locality, and hence criticality, is obtained by fine-tuning 
the control parameters, precisely as in continuous phase
transitions. In this paper we focus 
on the similarities and differences between SOC
and models with absorbing states.
The latter are relatively well-understood:
we can identify the order parameter, and know the static and 
dynamic scaling behavior in the neighborhood of the
critical point.  The mean-field and field-theoretic analysis are
well-established, and one has some idea how to derive these starting 
from an exact
master equation \cite{peliti}.  Applying these ideas, 
we arrive at a new understanding of SOC.

In Sec. II we review the models of interest: contact processes and sandpiles.
The formulation of a general theory of SOC is problematic
because of the nonlocal interactions implicitly present in these models.
For instance, field-theoretic analysis encounters difficulties
related with the singularity of the continuum limit \cite{diaz}. 
Moreover, it is in general not possible to treat simultaneously
the two timescales of avalanche propagation and external driving.
Sec. III describes how these
problems can be solved by a suitable ``regularization''
of the dynamics, which is local in space and time, and presents 
sandpile criticality as a kind of absorbing-state transition.  
The regularized dynamics readily lends itself to a continuum 
formulation, presented in Sec. IV.
This motivates,
in Sec. V, a study of sandpiles at the critical point, without boundaries or
driving, and of the pair contact process subject to a slow drive.  We present
preliminary simulation results of these systems.  
We summarize our new perspective
on SOC in Sec. VI.

\section{Contact Processes and Sandpiles}

\subsection{The contact process}

One of the simplest models showing an absorbing-state transition is the
contact process (CP) \cite{cp} (see Fig.~\ref{fig:cppcp}a). To each site of a 
$d-$dimensional lattice we assign a binary variable $\sigma_i=0,1$.
(Occupied sites are said to harbor a `particle.')
Occupied sites ($\sigma \!=\!1$) become empty ($\sigma \!=\! 0$) at unit rate,
while empty sites become occupied with rate $w\lambda$, where
$w$ is the fraction of occupied nearest neighbor (NN) sites. 
The vacuum state (all sites empty), is clearly absorbing and is
the only stationary state for $\lambda < \lambda_c$,
while for $\lambda > \lambda_c$ there is also an active stationary state.
($\lambda_c \simeq 3.298$ in one dimension.)
The order parameter is the density $\rho_a$ of active (occupied)  
sites, which vanishes at the transition as
\begin{equation}
\rho_a \sim (\lambda - \lambda_c)^\beta.
\end{equation}
The simplest (mean-field) description of the CP treats the $\sigma_i$ as
uncorrelated:
\begin{equation}
\frac{d \rho_a}{dt}  = - (1 - \lambda)\rho_a - \lambda \rho_a^2,
\end{equation}
leading to $\lambda_c = 1$ and $\beta = 1$.
As in equilibrium, we characterize the critical singularities
by a set of critical exponents \cite{torre}, such as $\beta$, and $\nu_{\perp}$,
 which describes the divergence of the correlation length $\xi$
\begin{equation}
\xi \sim (\lambda - \lambda_c)^{-\nu_{\perp}}.
\end{equation}
Besides the ``thermal" perturbation $\Delta \equiv \lambda - \lambda_c$,
a second relevant field is an external particle source $h$.
($h$ is the rate of ``spontaneous" creation at vacant sites.)
For $\Delta =0$, $\rho_a \sim h^{1/\delta_h}$.
Other exponents are defined by considering the decay of perturbations
to the stationary state \cite{torre}. 
Models with a single absorbing
state fall generically in the universality class of directed
percolation (DP), also known as Reggeon field theory \cite{torre,kinzel,cardy}.

A more complicated situation arises when many absorbing
configurations exist. 
The simplest model to have
been studied in detail so far is Jensen's pair contact process 
(PCP) \cite{pcp1} (see Fig.~\ref{fig:cppcp}b).  
In this model, a nearest-neighbor pair of 
particles may mutually annihilate, with probability $p$, or else,
with probability $q \equiv 1-p$, create a new particle at a randomly
chosen NN, provided it is vacant.
There are infinitely many absorbing configurations,
since all that is required is the absence of any NN particle pairs.
In one dimension the static critical behavior 
at $q_c = 0.9229$ is DP-like \cite{pcp1,MUNOZ}, but the
spreading or avalanche dynamics has variable exponents, depending on the
particle density $\phi$ in the environment of the seed \cite{pcp2}.
A special, ``natural'' class of absorbing configurations with particle 
density $\phi_{nat}$ are those spontaneously 
generated by the critical dynamics. DP spreading
exponents are recovered only if the initial particle 
density is set to $\phi_{nat} \simeq
0.242(1) $ \cite{pcp2,pcp3,jmendes,dick96}.

\subsection{The sandpile model}

Sandpile models are cellular automata (CA) with an integer
(or in some cases continuous), variable $z_i$ (``energy"), defined on 
a $d-$dimensional lattice. At each time step
an energy grain is added to a randomly chosen site, until
the energy of a site reaches a threshold $z_c$.
When this happens the site relaxes 
\begin{equation}
z_i\to z_i -z_c
\end{equation} 
and  energy is transferred
to the nearest neighbors 
\begin{equation}
z_j\to z_j +y_j.
\end{equation}
The relaxation of a site can induce NN sites to relax 
in turn, if they exceed the threshold because of the energy received,
and so on.  From the moment a site reaches threshold, until all
sites have again relaxed ($z_i < z_c, \; \forall i$), the addition of energy
is suspended.  The sequence of events during this
interval constitutes an {\em avalanche}.
For conservative models the
transferred energy equals the energy
lost by the relaxing site ($\sum y_j=z_c$), at least on average.
Usually, dissipation occurs only at
the boundary, from which energy can leave the system.

Since the energy input stops during an avalanche, we have, in effect,
an infinite timescale separation between the toppling dynamics and
the external source.
Under these conditions the system reaches a stationary
state characterized by avalanches whose sizes $s$ 
are distributed as a power law \cite{btw,kada,grasma,manna1,lubeck}
\begin{equation}
P(s)\sim s^{-\tau}.
\end{equation}
The model originally introduced by Bak, Tang and Wiesenfeld (BTW)\cite{btw} 
is a discrete  automaton in which $z_c=2d$ and $y_j=1$ 
(see Fig.~\ref{fig:btw}).

An interesting variation  of the original sandpile is the 
Manna model \cite{manna} (see Fig.~\ref{fig:manna}). 
In this automaton the critical threshold is $z_c=2$ independent of the 
dimensionality $d$, and  if a relaxation (toppling) takes place, the energy
is distributed to two randomly chosen nearest-neighbor sites
(see Fig.~\ref{fig:manna}). 
Variations in which part of the energy is 
kept by the relaxing site can also be considered, as well as  
models in which energy is transferred along 
a preferred direction \cite{kada}.

Finally, sandpile models in which part of the energy is dissipated have 
been studied \cite{diss}. 
In continuous-energy models, some fraction of the energy removed from
a relaxing site is lost, instead of being transferred to one of the 
neighbors \cite{diss}.
In a discrete-energy model, such as the Manna or BTW
sandpiles, one can introduce a parameter $\epsilon$
representing the average energy dissipated in an elementary relaxation event.
The two dissipation mechanisms lead to the same effect,
namely a characteristic length is
introduced into the system and criticality is lost.
The avalanche size distribution decays as
\begin{equation}
        P(s)\sim s^{-\tau}f(s/s_c)  ,
                                                \label{eq:D(s)}
\end{equation}
where the cutoff size scales as $s_c\sim \epsilon^{-1/\sigma}$.
We can also observe avalanches in the contact process, by starting the system
with a single particle \cite{paczuski96}.  
The activity may spread over many sites
before dying out; avalanches are power-law distributed if $\lambda $ 
is set to its critical value.

At first glance, the BTW sandpile looks quite unlike the CP.  One difference 
is that the avalanche dynamics in the sandpile is 
{\em nonlocal} and {\em deterministic}. 
The sandpile model is inherently nonlocal because of the implicit 
time scale separation. A site can receive energy only if the system is 
quescient, i.e., no active sites are present on the lattice.
This implies that transition rates depend upon the entire set 
of lattice variables present in the system, giving rise to a strongly 
nonlocal dynamical rule. Given
the configuration prior to the avalanche, and the location of the newly
added particle, deterministic toppling rules govern the evolution to the next
stable configuration, and this evolution can affect sites anywhere in 
the system.
To have any hope of applying the methods used for the CP, we have to assume
that the deterministic sandpile dynamics can be realized as 
a limiting case of models with local, stochastic dynamics, 
{\em belonging to the same universality class as the sandpile}.
(We refer to this as ``regularizing'' the sandpile rules.)
The latter hypothesis would need to be verified, but seems plausible
if the rules respect the same symmetries and conservation laws 
as those of the original
model.  (We provide an example in Sec.~V.)
Similar considerations apply to `extremal dynamics,' which
requires the action of an omniscient agent to choose the next
event.  Such a dynamics can presumably emerge as a limiting case of local rules
in which each unit only has information about a finite number of neighbors.

As originally defined, the sandpile seems to involve {\em no} parameters.  
There is only the
toppling rule, which, after some time, miraculously yields
a critical state. But in devising a regularized dynamics, we 
are forced to include a nonvanishing driving rate by  introducing the 
probability $h$ 
per unit time that a site will receive a grain of energy \cite{vz}. 
(We may fix the relaxation rate for active sites at unity.)  Energy
is distributed homogeneously and the total energy flux is given by
$J_{in}=hL^d$. The parameter $h$ 
sets the driving timescale or equivalently the typical
waiting time between different avalanches as $\tau_d \sim 1/h$.
As $h\to 0$, we recover the slow driving limit, i.e., during an 
avalanche the system does not receive 
energy. This formulation of the dynamics 
has the advantage of being local in space and time. The state 
of a single site depends only on the state of the site itself and its 
nearest-neighbor 
sites at the previous time step, through a transition probability
that is given by the reaction and driving rates. 

\section{Towards a local theory of SOC}

After reformulating the sandpile rules as local and stochastic,
we can proceed along the path followed for nonequilibrium
phase transitions. From the master equation 
we can derive mean-field equations that give a qualitative
picture of the phenomenon, exploiting several analogies with
models with absorbing states.
The mean-field analysis of regularized sandpiles shows that the order 
parameter is the density $\rho_a$ of active sites
(i.e., whose height $z \geq z_c$), and that 
$\rho_a$ is coupled to the densities of `critical' ($z=z_c - 1$) and 
`stable' ($z < z_c -1$) sites \cite{vz}.  In mean-field theory,
the dependence of the order parameter
$\rho_a$ on the parameters $h$ and $\epsilon$ can
be obtained on the back of an envelope.
Since energy is conserved in the stationary state, the incoming
energy flux $J_{in}$ must be balanced by the dissipated
energy $J_{out}=\epsilon\rho_a L^d$. From $J_{in}=J_{out}$, we
obtain
\begin{equation}
\rho_a=\frac{h}{\epsilon}.
\label{eq:rhoa}
\end{equation}
There is no stationary state for $h > \epsilon$; see Fig.~\ref{fig:phase}.
The model is critical just in the 
double limit $h,\epsilon \to 0, h/\epsilon\to 0$, since
the zero-field susceptibility $\chi\equiv d\rho_a/dh$ diverges, 
implying a long-ranged (critical) response function.
Critical behavior emerges in the 
limit of vanishing driving field, corresponding to locality-breaking 
in the sandpile dynamics.     
The driving and dissipation rates
are the {\em control parameters} of the model; the stationary 
order-parameter naturally vanishes at the critical point.
When $h=0$, any configuration with $\rho_a = 0$ is
absorbing.  Thus there are an infinite 
number of absorbing configurations for a
sandpile, just as for the
pair contact process.  
(In close analogy with Ref.~\cite{vz}, 
in the field theory of the PCP the order parameter --- the density of 
nearest-neighbor {\em pairs} ---
is coupled to a non-order-parameter field \cite{MUNOZ}.)

In absorbing-state transitions, it is very useful to
consider the spread of active sites from an isolated seed. 
Following the scaling framework developed by Grassberger and 
de la Torre \cite{torre}, we expect that
the probability that a small perturbation imposed 
on an absorbing configuration activates 
$s$ sites scales as
\begin{equation}
P(s,\epsilon)=s^{-\tau}{\cal G}(s/s_c(\epsilon)),
\end{equation}
where $s_c\sim\epsilon^{-1/\sigma}$ is the cutoff in the avalanche size. 
The perturbation decays in the  stationary subcritical state as 
\begin{equation}
\rho_a(t) \sim t^{\eta}{\cal F}(t/t_c(\epsilon)).
\end{equation}
Here $t_c$ denotes the characteristic time which scales as
$t_c \sim \epsilon^{-\nu_{||}}$.
In this way we have translated the avalanche description
into the formalism commonly employed to study models with 
absorbing states \cite{paczuski96}.

It is natural to regard sandpile models as 
having two parameters, $\epsilon$
and $h$, with the original models poised, 
by definition, at the point ($0,0^+$).
It should be evident that $\epsilon$ in the sandpile model is 
a `temperature-like'
variable, playing the same role as 
$\lambda_c - \lambda \equiv - \Delta $ in the CP.
For $\epsilon > 0$ we cannot have sustained avalanches; 
they decay exponentially
in this {\em subcritical} regime.  
To have self-sustained avalanches, or an active
stationary state with $h \equiv 0$ in the sandpile, 
we would need $\epsilon < 0$,
that is to say, the possibility of creating additional 
energy quanta when a site topples.
But this immediately raises a new problem: the 
energy will never be lost (except
at the boundaries), so in the thermodynamic limit
we shall have a runaway `chain reaction' instead of a
stationary state for $\epsilon < 0$!  The impossibility of a stationary 
state for $\epsilon < 0$
is analogous to the absence of a well-defined free energy 
in the Gaussian model below $T_c$.
Neither model has the saturation effect needed for stability 
in the `low-temperature' phase.
In the Gaussian model the stabilizing $u\phi^4$ 
term is missing from the Hamiltonian,
while in the sandpile there is nothing to stop energy accumulating.  
Indeed, to do so would mean to lose
energy from the system, destroying the conservation 
law even for $\epsilon = 0$.  
Criticality would then {\em require} that
$\epsilon $ take some negative but {\em a priori unknown} value.  
{\em Thus the possibility
of not having to tune the system is predicated 
on the absence of saturation}, or,
equivalently, on having strict energy conservation when $\epsilon = 0$.

Another example is furnished by removing saturation (i.e., the restriction
to at most one particle per site) from the CP, resulting 
in an exactly-soluble birth-and-death
branching process.  In this model, particles disappear at unit rate, 
and produce offspring at rate $\lambda$, at neighboring sites,  
whether they are occupied or not.
This corresponds to setting $b=0$
in the mean-field equation
\begin{equation}
\frac {d \rho}{d t} = (\lambda -1) \rho - b \rho^2.
\end{equation}
The density grows without limit for $\lambda > \lambda_c = 1$.  
It is important to
note that $\lambda_c = 1$ not only in mean field theory, 
but in fact for the actual
birth-and-death process.  Avalanches follow power laws, with the 
survival probability $P(t) \sim t^{-1}$, for example.  Restoring
saturation ($b > 0$) permits the existence 
of an active stationary state, but at the cost
of shifting $\lambda_c$ to some larger but 
{\em a priori unknown} value.
(The birth-and-death process is
free of higher-order terms that would renormalize the critical value of the
thermal parameter from that given by mean-field theory.) 
As shown in Fig.~\ref{fig:phase},
neither model has a stationary state for negative values of the thermal
parameter $r$; this is the main difference from the phase diagram of the CP.
The sandpile, however, presents a further subtlety: while there is no upper bound
on {\em energy}, the order parameter is subject to saturation, 
since $\rho_a $ cannot exceed unity!  

The critical point of the birth-and-death process is 
at $\lambda =1$ because this point
corresponds to a balance, on average, between births 
and deaths.   Similarly, 
the sandpile is critical at $\epsilon =0$ because a toppling 
site sends a particle to each of $g$ neighbors, and each of these
neighbors is critical with probability $1/g$,
so the gain and loss terms for the number of active sites
balance on average.  
Thus the sandpile and the birth-and-death 
process have the same phase diagram.  
This does not mean, of course, that the two models share
the same avalanche dynamics --- that of the 
birth-and-death process is rather trivial.
An important aspect of the sandpile is that the condition needed for
critical avalanches --- that a fraction $1/g$ of 
the nearest neighbors of an active site  
be critical --- is established by the transient
dynamics of the model.  The leftovers from preceding avalanches
provide the environment in which activity is critical.   
Memory appears to be the crucial
feature of SOC models, and is due to the presence of a nontrivial threshold for
activity; for sandpiles this means that $z_c \geq 2$.  (For completeness,
we note that a sandpile with $z_c = 1$ corresponds to a simple random walk,
with well-known scaling properties.  One may think of it as the analog of the 
birth-and-death process, in the family of models obeying strict conservation of
particle number for $\epsilon = 0$.)  
The mean-field analysis \cite{vz} shows that
having (on average) a fraction $1/g$ of the nearest 
neighbors of an active site  
critical is necessary for having a 
stationary state, in which energy input
is balanced by dissipation.  That is, the 
only stationary state for a sandpile at
$(0,0^+)$ is a critical state.

What happens when we impose an activity threshold on the CP?
One realization of such a threshold corresponds to the PCP.  We 
study the {\em driven} PCP in Sec. V.  
Here we consider the field-theoretic 
description of the PCP \cite{MUNOZ}.
As noted above, the order parameter $\rho$ is coupled to a 
non-order parameter field $n$ representing the density
of isolated particles.  The equations take the form
\begin{equation}
\frac {\partial \rho}{\partial t} = D_{\rho} \nabla^2 \rho - a\rho 
-b\rho^2 -wn\rho + \cdots + \eta_{\rho}
\label{pcprho}
\end{equation}
and
\begin{equation}
\frac {\partial n}{\partial t} = 
D_n \nabla^2 \rho  +r\rho 
-u\rho^2 -\overline{w}n\rho + \cdots + \eta_{n},
\label{pcpn}
\end{equation}
where the noise terms satisfy
\begin{equation}
\langle \eta_i (x,t) \eta_j (x',t') \rangle = 
\Gamma_{i,j} \rho (x,t) \delta(x-x') \delta (t-t').
\end{equation}
The field $n(x,t)$ is frozen in regions where $\rho = 0$.
(If $w=0$, Eq. (\ref{pcprho}) is the minimal field theory 
for the CP \cite{janssen}.)
Now, because of the simple form of the $n$ equation, 
we can formally eliminate this field to obtain
\begin{equation}
\frac {\partial \rho}{\partial t} = 
D_{\rho} \nabla^2 \rho - a\rho -b \rho^2 + \eta_{\rho}
-wr\rho (x,t) \int _0 ^t dt' \rho(x,t') 
e^{-\overline{w} \int _{t'} ^t ds \rho(x,s) }
\end{equation}

\noindent which exhibits a long-memory effect \cite{MUNOZ}.  The
nonlocal term turns out to be irrelevant to the stationary properties
of the active phase: it is exponentially small if the density
of active sites is different from zero. The situation can be different 
for spreading from a seed, in which case the active sites form
only an infinitesimal fraction of the lattice. 

\section{Field theory of sandpiles}

A field theory of sandpiles should parallel that for the PCP in many respects.
As noted before, a crucial point of the sandpile dynamics is 
the coupling of the density field $\rho_a$ with the background of
critical sites $\rho_c$. 
Each region devoid of active sites is frozen until such a
site is generated. The activity spreads 
and in general alters the configuration before it moves away or disappears. 
The active sites leave a trace of their dynamical history 
in the frozen configurations of critical and stable sites they produce. 
If new active sites are created in the same region at some 
later time, they will feel the effect of the active sites
present earlier in the region. This  
creates a long-range interaction in time and space among 
active sites. The range of this interaction  depends
on the characteristic timescale of the driving, because the 
fluctuations induced by $h$ destroy the memory effect. 
Close to the infinite timescale separation, the characteristic 
driving timescale diverges and the range of the 
nonlocal interaction extends to the entire system. 
This picture is valid also for the PCP, since in both systems the
response function diverges as $\rho_a$ approaches zero and 
the nonlocal term becomes more and more important as $\rho_a\to 0$. 
In sandpile models the density of active sites 
is proportional to the external field $h$ and
nonlocality is recovered in the limit $h\to 0$.

We turn now to a detailed continuum description of 
the BTW model.
Let $\rho_i ({\bf x},t)$ be the density of sites with height $i $ at  {\bf x}.
We note  that each site is subject to an input of 
energy due to three sources:
(1) The external field, $h$; (2) toppling of active sites at 
any of the four NN's: 
$(4-\epsilon)\rho_a$ where $\rho_a = \sum_{i \geq 4} \rho_i $ and $\epsilon$ is
the average energy dissipated; 
(3) a diffusion-like contribution: $(1-\frac{\epsilon}{4})\nabla^2 \rho_a$.
The diffusive term arises because a gradient in $\rho_a$ leads to a particle
flux: the excess in the mean number of particles arriving at {\bf x} 
from the left,
over those arriving from the right, 
is $j_x ({\bf x},t) = -(1-\frac{\epsilon}{4})\partial_x \rho_a$.  
The net inflow of particles at {\bf x} is 
therefore $-\nabla \cdot {\bf j} = (1-\frac{\epsilon}{4})\nabla^2 \rho_a$.
Applying these observations to the mean-field equations 
derived in Ref. \cite{vz},
we can write down the following set of continuum equations:
\begin{equation}
\frac{\partial \rho_i}{\partial t} = \rho_{i+4} + (\rho_{i-1} -\rho_i) 
\{ (4-\epsilon)[\rho_a + \frac{1}{4}\nabla^2 \rho_a] + h \}
+ \eta_{i+4}^T - \eta_i + \eta_{i-1} \;, 0 \leq i \leq 3
\label{ft03}
\end{equation}
and
\begin{equation}
\frac{\partial \rho_i}{\partial t} = -\rho_i + \rho_{i+4} + 
(\rho_{i-1} -\rho_i) 
\{ (4-\epsilon)[\rho_a + \frac{1}{4}\nabla^2 \rho_a] + h \}
-\eta_i^T + \eta_{i+4}^T - \eta_i + \eta_{i-1} \;, i \geq 4.
\label{ft4}
\end{equation}

\noindent (For $i=0$ of course, $\rho_{-1} $ and $\eta_{-1}$ are
identically zero.)  The terms $\eta_i$ represent noise 
arising due to fluctuations in the 
number of events of a given kind; $\eta_i$ is the contribution
associated with the reaction $i \rightarrow i+1$,
and $\eta_i^T $ with toppling: $i \rightarrow i-4$ for $i \geq 4$.
Since the number of events is Poisson-distributed (approaching a Gaussian
in the continuum limit), the variance equals the mean, 
and the noise variance is
proportional to the mean rate of the corresponding process.  Thus we have 

\begin{equation}
\langle \eta_i ({\bf x},t) \eta_j ({\bf x'}, t') \rangle = 
\Gamma \delta_{ij} \delta({\bf x}-{\bf x'})
\delta(t-t') \rho_i ({\bf x},t) [ (4-\epsilon)\rho_a ({\bf x},t) + h] 
\label{covij}
\end{equation}

\noindent and

\begin{equation}
\langle \eta_i^T ({\bf x},t) \eta_j^T ({\bf x'}, t') \rangle 
= \Gamma^T \delta_{ij} \delta({\bf x}-{\bf x'})
          \delta(t-t') \rho_i ({\bf x},t) \;.
\label{covTT}
\end{equation}

\noindent The noise terms $\propto \nabla^2 \rho_a$ 
have been dropped, as they are expected to be irrelevant.

This set of equations satisfies probability conservation: 
$\sum_s \rho_s $ is a constant,
equal to unity by normalization.  
Let $\zeta ({\bf x},t) \equiv \sum_s  s \rho_s ({\bf x},t) $ 
be the local energy density.
{}From Eqs. (\ref{ft03}) and (\ref{ft4}) we have
\begin{equation}
\frac{\partial \zeta}{\partial t} = 
(1-\frac{\epsilon}{4})\nabla^2 \rho_a +  h  
- \epsilon \rho_a + \eta_{\zeta} \;,
\label{dzetadt}
\end{equation}
where
\begin{equation}
\langle \eta_{\zeta} ({\bf x},t) \eta_{\zeta} ({\bf x'}, t') \rangle = 
\Gamma^{\zeta} 
          \delta({\bf x}-{\bf x'}) \delta(t-t') 
[h + \epsilon \rho_a ({\bf x},t)] \;.
\label{covzeta}
\end{equation}
(Again we have neglected diffusive noise, 
and have used the fact that in the absence
of a source and of dissipation, the total energy does not fluctuate.)  For
$h=\epsilon=0$ we have simply
\begin{equation}
\frac{\partial \zeta}{\partial t} = \nabla^2 \rho_a ,
\label{dzeta00}
\end{equation}
so that $E = \int d^2 x \zeta ({\bf x},t) $ is conserved.

The generalization of Eqs. (\ref{ft03}) 
and (\ref{ft4}) to other dimensions, or to
other sandpile models (e.g., Manna's), is straightforward, but
the analysis of this complicated set of equations
is problematic.
One might try to cut off the hierarchy by simply declaring $\rho_i \equiv 0$
for $i$ greater than some $i_c$.  The choice of cutoff, however, 
is not obvious, and one would have to add 
suitable correction terms to ensure that
energy is conserved when one active site topples onto another.
(Altering the sandpile rules to {\em forbid} such transfers --- in effect,
constraining all sites to have $z \leq z_c$ --- raises an interesting
possibility, but one that we shall not pursue further here.)
As a step toward simplification of the continuum equations, we
sum up Eq. (\ref{ft4}) for $i \geq 4$ to obtain
\begin{equation}
\frac{\partial \rho_a}{\partial t} = -\rho_a + \rho_a^* + \rho_3 
\{ (4-\epsilon)[\rho_a + \frac{1}{4}\nabla^2 \rho_a] + h \}
-\eta_a^T + \eta_a^* + \eta_3 \;,
\label{fta}
\end{equation}
where $\rho_a^* \equiv \sum_{i\geq 8} \rho_i$, 
$\eta_a^T \equiv \sum_{i \geq 4} \eta_i^T$,
and $\eta_a^* \equiv \sum_{i \geq 8} \eta_i^T$.  
Since the density of sites with
heights $\geq 8$ should be negligible, 
we might ignore the starred terms.  Then the
active-site density is coupled only to $\rho_3$, 
identified in Ref. \cite{vz} as the
density of {\em critical} sites, $\rho_c$.  In that work, sites
with heights $< 3$ are considered in a unified manner, as the 
density of {\em stable} sites, $\rho_s$.  Eq. (\ref{ft03}) shows,
however, that the evolution of $\rho_c$ is coupled 
specifically to $\rho_2$, not simply
to $\rho_s \equiv \rho_0 + \rho_1 + \rho_2$.  
In the mean-field theory \cite{vz}, the
quantity $u= \rho_2/\rho_s$ is therefore introduced.  
In a spatially homogeneous
stationary state, the value of $u$ can be deduced 
from energy conservation.  But
in the present context, $u = u({\bf x},t)$ 
is another dynamical variable.  Thus 
our attempt to reduce Eqs. (\ref{ft03}) and (\ref{ft4}) 
to a description in terms of
three basic categories meets with difficulties.

Rather than pursuing a systematic derivation of a 
reduced set of equations from
Eqs. (\ref{ft03}) and (\ref{ft4}), we shall use what we have 
learnt so far, together
with the observation that in constructing a field theory, a detailed accounting
is unimportant, so long as one respects the symmetries and conservation laws
of the original model.  In the present instance, it is essential 
to ensure conservation of energy when $\epsilon =h =0$.  
In fact, Eq. (\ref{dzetadt})
represents this explicitly, and shows how 
the energy density $\zeta$ is coupled 
to $\rho_a$.  We therefore retain Eq. (\ref{dzetadt}) as 
one of our basic equations.

We obtain the other equation by replacing 
$\rho_3$ in Eq. (\ref{fta}) with $f(\zeta) (1-\rho_a)$: only non-active 
sites can contribute to
the gain term for $\rho_a$, and they do so at a rate that depends on the
local energy density.  
($f(\zeta)$ plays a role analogous to that of $u$ in the mean-field theory.)
For small $\zeta$, far from criticality, 
one expects the height distribution to be
Poissonian, so that $f(\zeta) \propto \zeta^3$; for large values of $\zeta$,
$f$ will be a decreasing function of $\zeta$.  
The case of most immediate interest is a system near the 
critical stationary state, with 
$\zeta \simeq \zeta_c$ and $f \simeq \rho_c$, where $\zeta_c$ 
and $\rho_c$ represent
the average values of energy and the 
density of critical sites, respectively, at the critical point.
{}From the MF solution  we have $\rho_c=1/4$ for the BTW 
model in the limit $h\to0$, 
i.e., the density of critical sites is still rather small, and 
$f$ is an increasing
function: $f(\zeta) = \rho_c + A (\zeta -\zeta_c) + \cdots$, with $A>0$.  
Presumably, only the linear term need be retained 
in the vicinity of the critical point.  
The resulting
field theory for the regularized BTW sandpile is given by
\begin{equation}
\frac{\partial \rho_a}{\partial t} = -\rho_a  + [\rho_c  +A(\zeta -\zeta_c)](1-\rho_a)
\{ (4-\epsilon)[\rho_a + \frac{1}{4}\nabla^2 \rho_a] + h \}
                                              +\eta_a \;,
\label{fta1}
\end{equation}
where
\begin{equation}
\langle \eta_a ({\bf x},t) \eta_a ({\bf x'}, t') \rangle = 
\Gamma \delta({\bf x}-{\bf x'})
          \delta(t-t') \zeta({\bf x},t) 
[ (4-\epsilon)\rho_a ({\bf x},t) + h] \;,
\label{covaa}
\end{equation}
together with  Eqs. (\ref{dzetadt}) and (\ref{covzeta}).
As in the PCP, our field theory for the sandpile supports an infinite number
of absorbing configurations: any $\zeta({\bf x},t)$ consistent with
$\rho_a \equiv 0$ (when $h=0$). 
In both theories, the non-order-parameter field enters the
equation for the order parameter in the role of an 
{\em effective creation rate}.
The crucial difference between the PCP and the sandpile is that in the latter
case, this auxiliary field is conserved at the critical point.  

In a simple
mean-field treatment (spatially homogeneous, no noise), we have
\begin{equation}
\frac{d\rho_a}{dt} = -[\frac{\epsilon}{4} 
- a(\overline{\zeta}-1)(1- \frac{\epsilon}{4})] \rho_a 
- [1 + a(\overline{\zeta}-1)](1-\frac{\epsilon}{4}) \rho_a^2
                           + [1 + a(\overline{\zeta}-1)]\frac{h}{4},
\label{ftamf}
\end{equation}
and
\begin{equation}
\frac{d\zeta}{dt} = -\epsilon \rho_a    + h,
\label{ftzmf}
\end{equation}
where we define $4A(\zeta -\zeta_c)= 
a(\overline{\zeta}-1)$ by introducing 
$\overline{\zeta} \equiv \zeta/\zeta_c$.
For $\epsilon$ and $h$ small, and $h/\epsilon <<1$, the mean-field equations
have the stable stationary solution $\rho_a = h/\epsilon$, 
$\overline{\zeta} = 1-h/a\epsilon$.
In the case $h=0^+$ --- 
the slowly-driven limit $h, \epsilon \rightarrow 0$, 
with $h/\epsilon \rightarrow 0$ --- 
the stationary value of $\zeta$ approaches
the critical height $\zeta_c$. It is easy to recognize then 
that $\epsilon$ plays the role of a control parameter,
analogous to $\lambda$ in the CP, with the
critical point at $\epsilon=0$.  

A different situation is faced when we impose  $\epsilon=h=0$  
from the outset, rather than via the slow-driving limit. 
We have from Eq.(\ref{ftzmf}) that $\zeta$ is 
strictly conserved in this case. The average energy density 
is thus an external parameter that can 
be freely fixed in the initial condition. In this case, in fact, 
$\zeta$ is the only 
control parameter. In the following section we present 
simulations of just such a situation.

The full analysis of the field theory will be 
deferred to a future publication.  Here we
simply observe that for $\epsilon=h=0$, 
\begin{equation}
\zeta({\bf x},t) = \zeta({\bf x}, t=0) + \int_0^t dt' \nabla^2 
\rho_a ({\bf x},t') .
\label{zetat}
\end{equation}
The evolution of the active-site density contains long-memory terms.
If $\zeta ({\bf x},t=0) = \zeta_0 ({\bf x}) \approx \zeta_c$, then to leading
order
\begin{equation}
\frac{\partial \rho_a}{\partial t} =  \frac{1}{4} \nabla^2 \rho_a 
   +a (\overline{\zeta}_0 -1)\rho_a  
 - \rho_a^2 + \frac {a}{\zeta_c} \rho_a \int_0^t dt' 
\nabla^2 \rho_a ({\bf x},t')
                +\eta_a \;,
\label{fta2}
\end{equation}
Unlike the PCP, in which the memory terms decay 
$\propto \exp[-C\int dt \rho_a]$,
here the memory decays more slowly, via the diffusive relaxation of $\rho_a$.
It is worth noting that even in active regions, 
fluctuations in the height field
$\zeta$ cannot relax directly; they only do so by inducing similar 
fluctuations in $\rho_a$.  
Relaxation of the latter then redistributes energy along with active sites.

In summary, all of the models discussed 
so far --- sandpiles, the CP, PCP,
and the birth-and-death process --- have a critical point in a space of
two relevant parameters, one temperature-like ($r$), the other field-like ($h$).
Criticality requires $h=0$. 
Models such as the sandpile and PCP have a nontrivial 
threshold for activity and therefore exhibit multiple absorbing configurations.
When such models are run at ($r=0,h=0^+$), then
out of a range of possible values for one 
or more non-order-parameter
densities (the critical-site density in the sandpile, the density of isolated
particles in the PCP), the dynamics selects a unique value.
{}From this vantage, the fact that certain models are
critical by definition is
of secondary importance.  The essential feature is the behavior of a
critical system under slow drive.  We can study a critical, but non-SOC
sandpile by setting $\epsilon=h=0$; conversely, we can observe
avalanches on all scales in the PCP if we set $p=p_c$
and $h=0^+$.

\section{Simulation Results}

The preceding discussion motivates several new kinds of simulations of
the sandpile and the PCP.  We report some preliminary results in this
section.

\subsection{Sandpiles at $\epsilon = h = 0$}

In a regularized theory of sandpiles, we need to introduce a
dissipation rate $\epsilon \geq h$ to realize a stationary state.  
In the original model, a stationary state is achieved
by imposing open boundary conditions. 
While this may be appropriate for modeling processes in which stress may only
be released at the boundaries of the system, it is an inconvenience 
theoretically:
it is easier to study criticality in uniform systems; 
once bulk behavior is understood,
the effects of various kinds of boundaries can be analyzed.  We therefore
study a sandpile with periodic boundaries.
We performed simulations of the stochastic BTW sandpile at
$(\epsilon=0, h=0)$.
With $\epsilon = h = 0$, the mean-height 
$\zeta \equiv N/L^d$ is strictly conserved;
it is an additional parameter at our disposal.

Initial configurations are generated by distributing at random a fixed number
$N$ of particles amongst $L^d$ lattice sites.  
(Since the initial configuration is on average spatially homogeneous,
all average properties such as densities and correlation functions are
translation-invariant.)  Once all $N$ particles
have been placed (but not before), the dynamics begins:
each active site (i.e., having $z \geq z_c = 2d$) topples
at unit rate.  In practice, we maintain a list of the current set
of $N_a$ active sites,
choose one at random as the next to topple, and update the list
following the redistribution of energy to the $2d$ neighbors.
At each toppling event, time is incremented by $1/N_a$ --- the mean
waiting-time to the next event.  We compute average properties
over a set of $N_{samp}$ independent
trials, each using a distinct initial configuration.
($N_{samp} = 10^3 - 10^5$ depending on the
lattice size and the distance from criticality.)

Sustained activity depends upon two factors.  First, there must be 
at least one active site in the initial configuration.  This
condition is trivially satisfied on large lattices, 
as the probability of having {\em no} active sites becomes
exponentially small.  (For large $L$, the initial height at a given site
is essentially a Poisson random variable, $P_n \simeq \zeta^n e^{-\zeta}/ n!$,
so the probability of having no active sites $\sim (1-P_{2d})^{L^d} $.)
The second requirement is that there should be
on average at least one
critical site amongst the nearest-neighbors of an active site.
One expects the latter condition to depend
sensitively on $\zeta$, raising the possibility of a phase transition
as we vary this parameter.

In one dimension, not surprisingly, we observe a rather simple
behavior.  For $N < L$, all trials die out rapidly, so that the
only stationary state is the vacuum.  For $N \geq L$, on the other
hand, virtually all trials survive indefinitely.  (We verified
this up to $L=1000$.  In some instances the system evolves to
a configuration of the form ...11112011111... in which the
active site must forever circulate.)  Thus we see a first-order
transition at $\zeta=1$; the stationary
active-site $\overline{\rho_a}$ density jumps from zero to about 0.15.

In two dimensions the non-driven sandpile exhibits a critical point.
Fig.~\ref{fig:rhoact} shows that the active-site density 
in surviving trials exhibits a
non-monotonic approach to its stationary value.  
By performing studies of this kind,
always being careful to check that the system has reached a stationary state,
we determined $\overline{\rho_a} (\zeta,L)$
for a range of $\zeta$ values and for $L$ = 20, 40, 80 and 160.
In Fig.~\ref{fig:beta}  we see that
$\overline{\rho_a}$ appears to increase continuously from zero
at a critical value of $\zeta$.  To fix $\zeta_c$ we study the
dependence of $\overline{\rho_a}$ on $L$, as it should follow a nontrivial
power-law ($\overline{\rho_a} \sim L^{-\beta/\nu_{\perp}}$ 
in the usual notation),
only at the critical point.  Fig.~\ref{fig:rhovl} 
shows $\overline{\rho_a}$ fall
owing a power-law
for $\zeta = 2.125$, but clearly not for 2.124 or 2.126, 
allowing us to conclude
that $\zeta_c = 2.1250(5)$ for the two-dimensional sandpile.  Indeed, 
this value for the mean height is in perfect agreement
with the exact result $\zeta = 2.1248...$ derived  
by Priezzhev for the {\em driven} sandpile \cite{prizz}.
We also verify that at $\zeta = \zeta_c$, each active site has, on average, one
critical nearest neighbor.  The overall density of critical 
sites is $\overline{\rho_c} = 0.434$, again in agreement with 
driven sandpile simulations \cite{katori}  
(At the critical point, about 10\% of
critical sites have heights in excess of 4.)

Having located the critical point, we can examine the critical scaling of
various quantities.  Fig.~\ref{fig:beta} shows a clear 
power-law dependence of the
active-site density on the distance from the critical point: 
$\overline{\rho_a} \sim (\zeta - \zeta_c)^{\beta} $ with $\beta = 0.59(1)$.
The dependence of on $\overline{\rho_a} (\zeta_c,L) $ on system size yields
$\beta/\nu_{\perp} = 0.67(1)$.   (Figures in parentheses denote two
standard deviations in a least-squares linear fit.)  We also monitored 
$P(t)$, the fraction of surviving trials at time $t$.  (Approximately half
of the trials appear to survive indefinitely at $\zeta_c$.)  Associated with
the (approximately exponential) approach of $P(t)$ to its limit is a
relaxation time, $\tau$.  We find that $\tau$ has a power-law dependence
on $L$ at the critical point: $\tau \sim L^{-\nu_{||}/\nu_{\perp}}$,
with $\nu_{||}/\nu_{\perp} = 1.86(8)$.  For comparison, we note the
values for DP in 2+1 dimensions: $\beta \simeq 0.58$, 
$\beta/\nu_{\perp} \simeq 0.80$, and $\nu_{||}/\nu_{\perp} \simeq 1.76$.
The similarity in $\beta$ values is curious, but the differences
in the other ratios indicate that the sandpile is not in the DP
universality class.  (This is as expected, given the differences between the 
sandpile and the CP discussed in Sec. III.)  Studies of correlation functions,
that will allow determination of $\nu_{||}$ and $\nu_{\perp} $ separately,
will be reported in a future publication.

In the simulations just described, we have fixed $\zeta$, one of the variables
that the dynamics selects in {\em driven} 
sandpiles with dissipation at the open boundaries.
We observe criticality just at the value $\zeta_c$ observed in the
driven case, and other variables such as the critical site density assume
the same value in the two cases.  In effect, we are able to study
sandpiles in either of two ``ensembles," one with fixed energy,
the other with this variable adjusted by the system dynamics.  
Open boundaries, which served, in earlier sandpile simulations,
as an outlet for accumulated energy, are now seen not to be
essential for criticality.
An independent study of energy-constrained sandpiles 
(again with periodic boundaries), confirms that avalanches
follow the same power laws as in the original BTW model \cite{chessa}.
Finally, we note that our observation of criticality --- at the same $\zeta_c$
as in the BTW model --- in a stochastic sandpile with fully local rules,
supports the expectation voiced in Sec. II, that we can study SOC using a
regularized dynamics.

\subsection{Driven Pair Contact Process}

The one-dimensional PCP has a continuous 
absorbing-state transition at $q_c$;
below this value of the creation probability, the system falls into one of
an exponentially large (with $L$) number of absorbing configurations, each devoid of
NN pairs.  In contrast with previous studies, here we study a
{\em driven} PCP.  Starting from an empty 
lattice, we add particles at
randomly chosen vacant sites, until a NN pair is formed.  We then suspend
the addition of particles, and permit the system dynamics, 
as described in Sec. II,
to operate, until the system again falls into an absorbing configuration.
We simulate a system of size $L=1000$ with periodic boundary conditions
and study the avalanche distributions for different values of $q$, with both
parallel and sequential updating. 

\begin{figure}[bt]
\caption{The density $\phi$ of isolated particles in the slowly driven, 
        one-dimensional PCP at $q_c$,
        approaches the natural value.}
\label{fig:pcp2}
\end{figure}
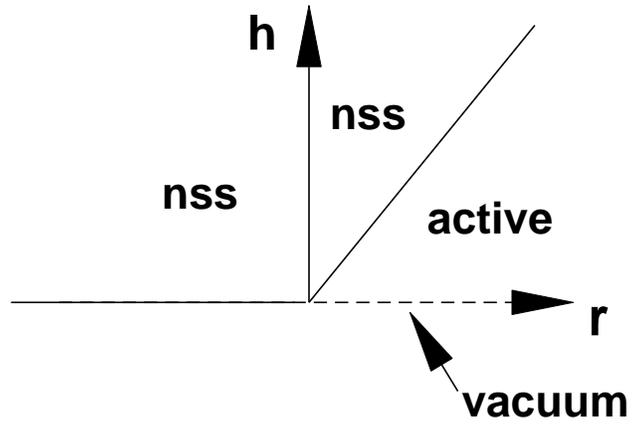
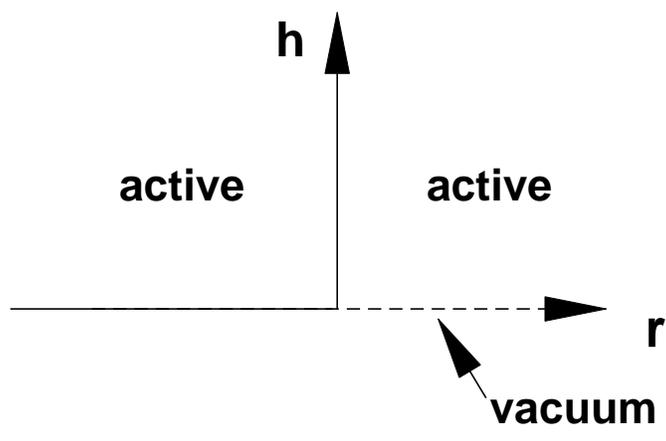
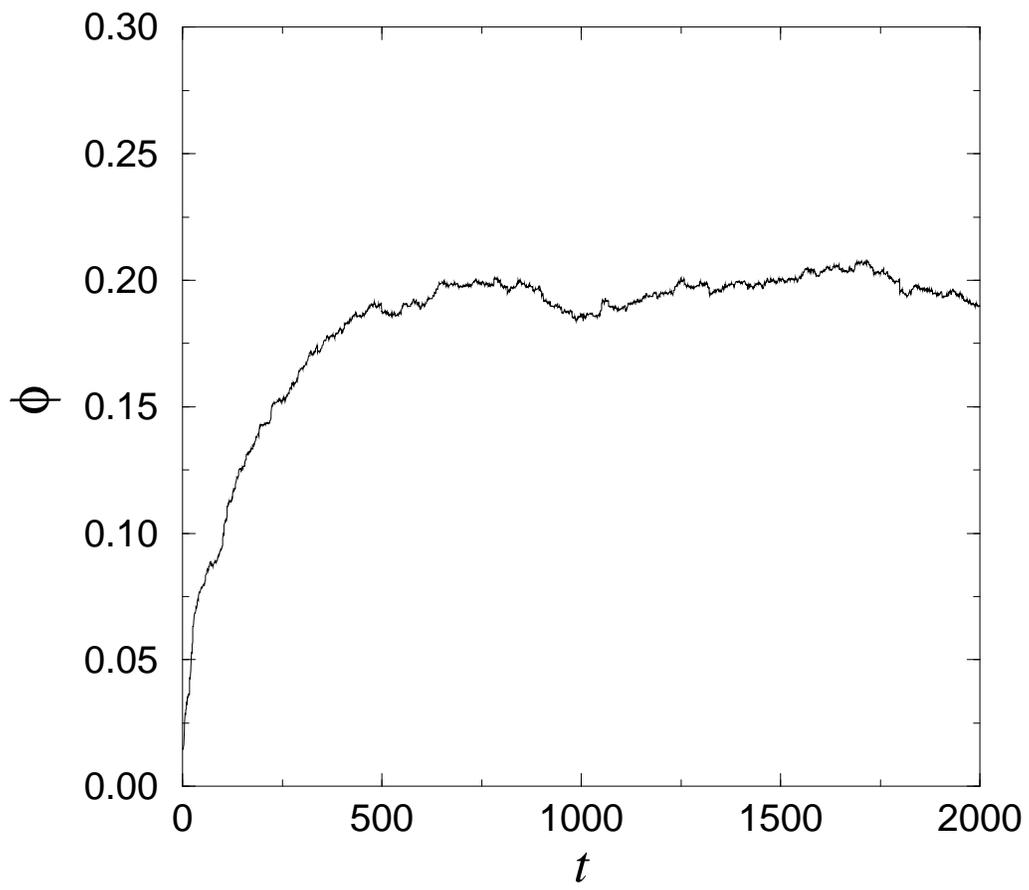
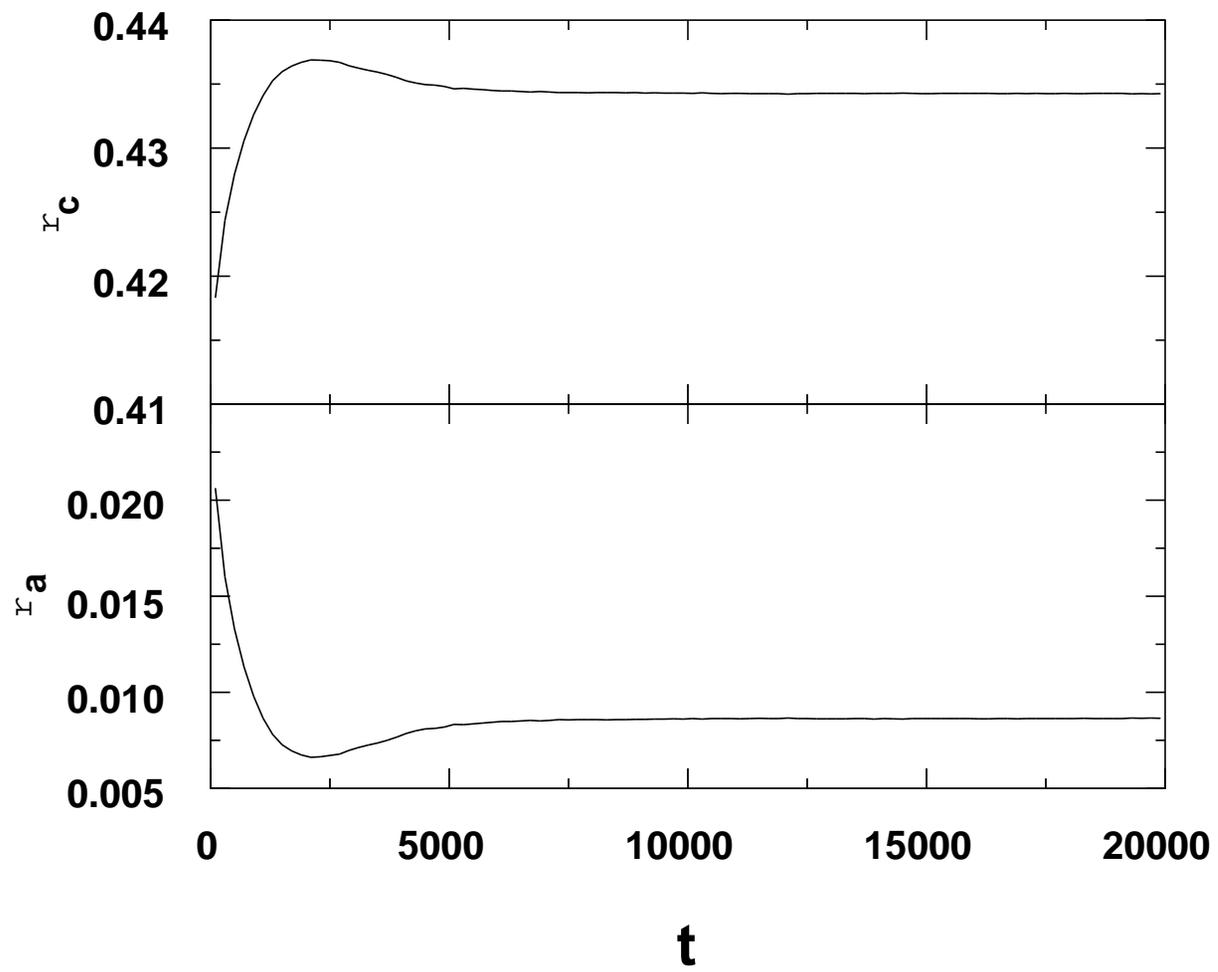
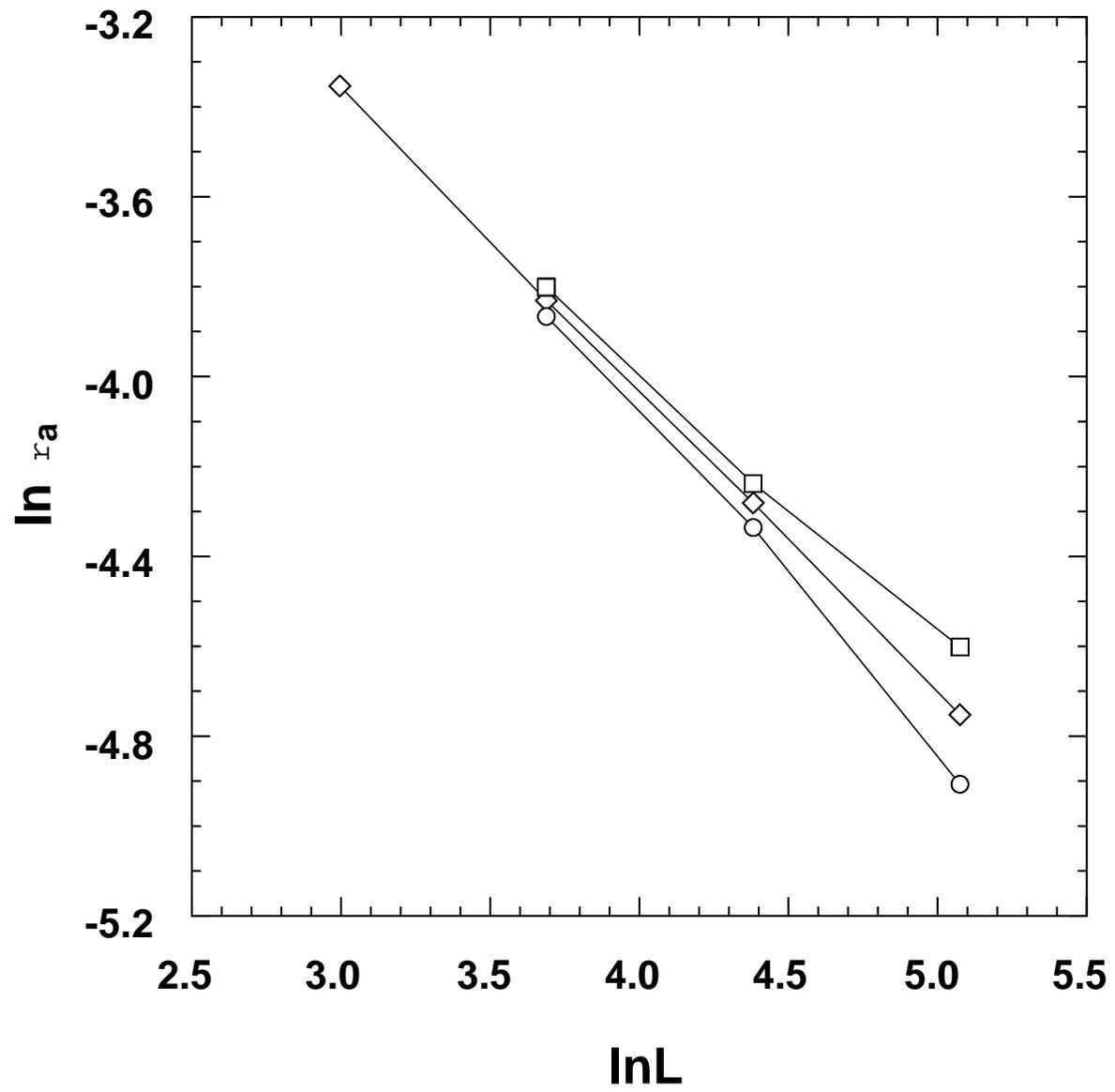
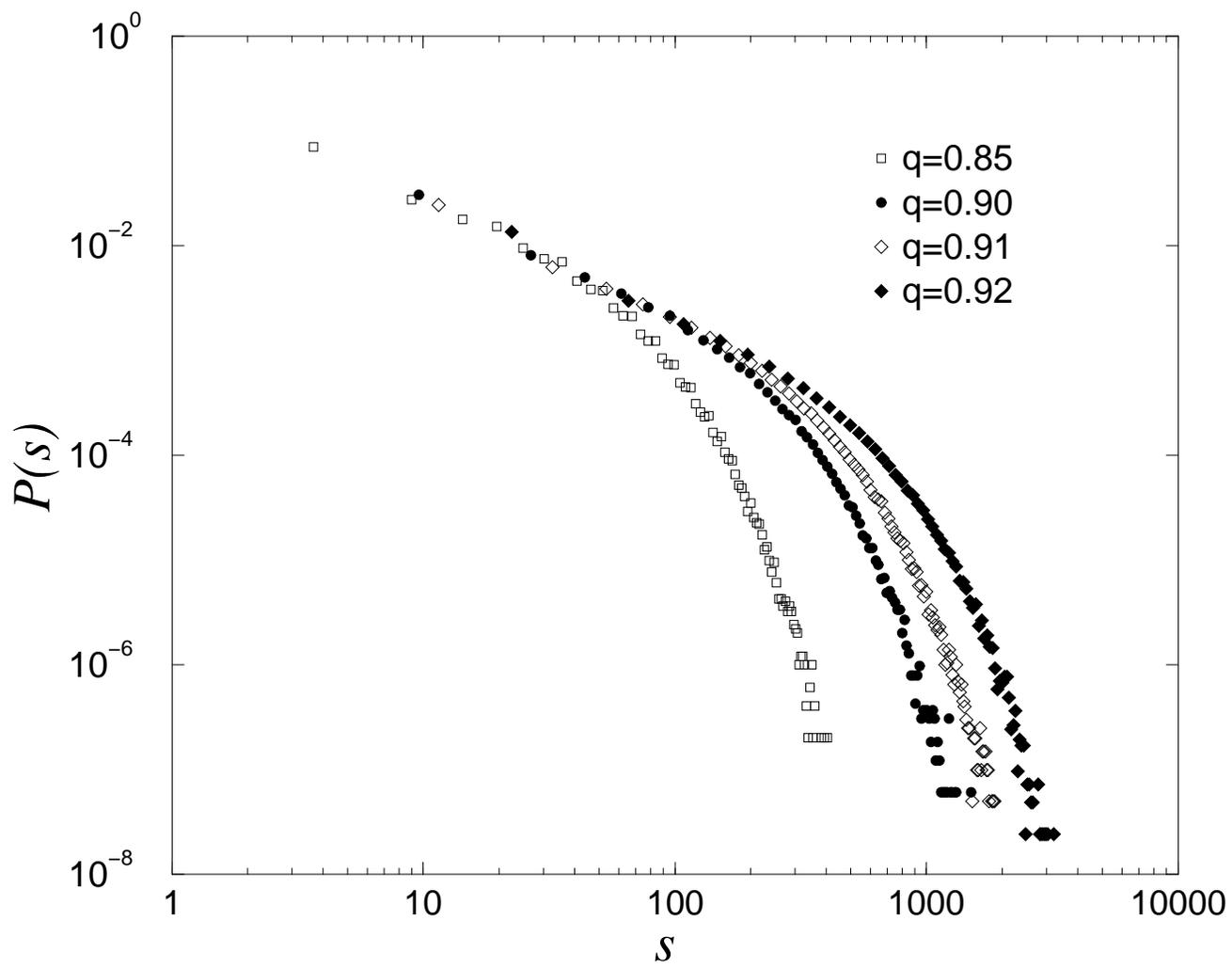

We collect statistics on the size and duration of the avalanches for various
values of $q$.  As illustrated in Fig.~\ref{fig:pcp1}, 
the avalanche-size distribution
$P(s)$ is power-law for some range of $s$, but suffers an exponential
cutoff at $s_c$, which grows as $q \rightarrow q_c$ as
\begin{equation}
s_c \sim (q_c-q)^{-1/\sigma}.  
\end{equation}
(Note that due to parallel updating, the critical creation rate
$q_c \simeq 0.95$ rather than 0.9229 as found in sequentially-updated
simulations.)  We see that the slope
of the power-law distribution is consistent with DP (i.e. $\tau = 1.08$).
Sequentially-updated simulations (not shown) yield $\tau =1.12$ and
$\sigma = 0.45$, very close to the expected DP value of 0.44.
In addition, we observe that at the critical point, the isolated-particle 
density approaches its natural value,
$\phi_{nat} \simeq 0.2$ (parallel updating) (see Fig.~\ref{fig:pcp2}).  
(Similarly, in the sequentially-updated case we observe 
$\phi \rightarrow \phi_{nat} \simeq 0.242$.)  A detailed comparison of
avalanche scaling under parallel and sequential dynamics will be 
presented elsewhere \cite{zvd}.

In the slowly-driven PCP, the system dynamics `self-organizes' 
the isolated-particle density $\phi$ to its natural value, the same as in the 
non-driven system.  This is similar to what happens in the sandpile, where the 
driven system selects the same
critical mean height, $\zeta_c$ that we found in simulations without driving.
There is, however, one rather striking difference between the models.  In
the PCP, activity can
spread at $q_c$ for {\em any} $\phi$ in the range [0,1/2].  
In the sandpile, by contrast,
activity cannot spread at all if the critical-site density 
is too low.  Each toppling
destroys an active site, and at least one of the neighbors 
must take its place
if activity is to persist.  In the PCP, each particle 
creation generates at least
one new pair as well, so the activity has a possibility 
of surviving even in an {\em empty}
lattice.  This suggests that one investigate a 
modified PCP, in which a pair
creates at particle at a (vacant) second neighbor, 
rather than at a NN; in this case
new pairs will only be formed if $\phi$ is sufficiently large.  
Other potentially interesting models are 
a saturation-free version of the PCP, 
and the PCP in two dimensions, where only two distinct 
universality class are predicted, namely DP and dynamical 
percolation \cite{miguel}. 
We defer investigation of these models to future work.

\section{Summary and perspective}

In this paper we have argued that SOC can be understood as an aspect
of multiple absorbing-state models under a slow drive.  We pointed out the
similarities in the phase diagrams of the two classes of models
(for the sandpile and the birth-and-death process, they are identical),
and in terms of avalanches and of bulk critical behavior, without boundary
dissipation.  We demonstrated that the sandpile exhibits an absorbing-state 
transition as we vary the mean height, and that the PCP, heretofore studied
only as an absorbing-state transition, exhibits a power-law avalanche
distribution under a slow drive.
We also suggested several new models to investigate, and derived a 
field theory of sandpiles.

Beyond these and other avenues for quantitative investigation, 
we propose a new viewpoint of SOC itself.  What `goes critical' in sandpiles
is $\rho_a$, the density of active sites.  
The evolution of $\rho_a$ is intertwined 
with other fields, which are frozen when $\rho_a = 0$.
These fields 
describe an energy density that is strictly conserved at the critical point.  
In order for avalanches to be critical,
two conditions are needed.  First, the parameters $h$ and 
$\epsilon$ must be set
to their critical values, i.e., to zero.  This is 
accomplished by the definition of the model, 
rather than by tuning parameters, but seems very similar 
in principle to criticality
in CP-like models.  The second condition is  that the environmental 
density is such as
to support avalanches on all scales.  Particle conservation
plays an essential role in this aspect, with 
the threshold for toppling providing for
a certain independence between $\rho_a$ and the overall particle density.
From this vantage, SOC is an absorbing-state transition riding
atop a substrate that preserves a record of the previous activity.  
SOC typifies the behavior under slow drive, at the critical point of a model
with an infinite number of absorbing configurations.

Finally, we offer a comment on the significance of sandpiles
as models or paradigms of physical processes.  The intention of the 
remark that the sandpile sits, by definition, at the critical point
in a two-dimensional parameter space, is not to trivialize 
it, but rather provide insight and access to new conceptual
and computational tools.  One may argue whether there is any
point introducing $\epsilon$ and $h$ as parameters for the sandpile; we merely
posit that their discussion seems natural if one wishes to draw an analogy
between sandpiles and other models with critical absorbing-state transitions.
The question``Why is Nature filled with systems that tune themselves to a
critical point?" may be replaced with: ``Why do so many systems share the
typical features of conservative, saturation-free 
dynamics, a threshold for activity, and widely
separated timescales for external driving, on one hand, and above-threshold
dynamics on the other."  The question of how this facet of Nature works remains
a deep one.

\newpage

\section*{Acknowledgements}
We are grateful to M. A. Mu\~{n}oz for helpful discussions and 
a careful reading of the manuscript.
We thank the ICTP, where this work was initiated, during the workshop ``The
Dynamics of Complexity''. The Center for Polymer Studies is supported by the NSF.
\vspace{2em}

\noindent$^*${\small Electronic address: dickman@lcvax.lehman.cuny.edu \\
Address as of 1 Jan. 1998: {\em Departamento de F\'{\i}sica, 
Universidade Federal de Santa Catarina, Campus Universit\'ario --- Trindade, 
CEP 88040-900, 
Florian\'opolis --- SC, Brazil}} \\
$^{\dagger}${\small Electronic address: alexv@ictp.trieste.it} \\
$^{\ddagger}${\small Electronic address: zapperi@miranda.bu.edu} \\

\begin{figure}[bt]
\caption{a) Transition rates in the one-dimensional contact process.  
Filled circles denote
occupied sites, open circles, vacant sites; gray sites may be either occupied
or vacant. b) Transition rates in the one-dimensional pair contact process.}
\label{fig:cppcp}
\end{figure}

\begin{figure}[bt]
\caption{The BTW sandpile model. When four grains are accumulated
        in one lattice site ($z_c=4$), the site relaxes distributing
        the grains to the neighboring sites.}
\label{fig:btw}
\end{figure}

\begin{figure}[bt]
\caption{The Manna sandpile model. When two grains are accumulated
        in one lattice site ($z_c=2$), the site relaxes distributing
        the grains to two randomly chosen neighbors.
        } 
\label{fig:manna}
\end{figure}

\begin{figure}[bt]
\caption{Phase diagrams of the sandpile and birth-and-death processes, 
and of the contact process and the PCP.  
The thermal parameter $r$ corresponds to $\epsilon$
in the sandpile, $1-\lambda$ in the birth-and-death process, and to 
$\lambda_c - \lambda$ in the contact process. `nss' denotes a region where
            no stationary state is possible.}
\label{fig:phase}
\end{figure}

\begin{figure}[bt]
\caption{Evolution of the active-site density $\rho_a$, and of the density
         $\rho_c$ of critical sites, in the two-dimensional
         stochastic sandpile at $\epsilon=h=0$.  System size $L=160$; 
        mean height $\zeta = \zeta_c = 2.125$.}
\label{fig:rhoact}
\end{figure}

\begin{figure}[bt]
\caption{Stationary active-site density 
        $\overline{\rho_a}$ in the two-dimensional
        stochastic sandpile at $\epsilon=h=0$, as a 
        function of $r \equiv \zeta - \zeta_c$.   
        The inset shows $\overline{\rho_a}$  versus mean height 
        $\zeta$ on linear scales.
        $+$: $L=40$; $\times$: $L=80$; $\diamond$: $L=160$.}
\label{fig:beta}
\end{figure}

\begin{figure}[bt]
\caption{Stationary active-site density $\overline{\rho_a}$ in 
        the two-dimensional
        stochastic sandpile at $\epsilon=h=0$, as a function of 
        system size $L$.   
        $\circ$: $\zeta =2.124$; $\diamond$: $\zeta=2.125$; 
        $\Box$: $\zeta=2.126$.}
\label{fig:rhovl}
\end{figure}

\begin{figure}[bt]
\caption{Avalanche-size distribution $P(s)$ in the slowly driven, 
        one-dimensional PCP, for various values of the creation 
        probability $q$. The system size is
        $L=1000$ and $10^6$ avalanches are recorded for each curve.}
\label{fig:pcp1}
\end{figure}

\end{document}